\documentclass[showpacs,showkeys,preprintnumbers,amsmath,amssymb,superscriptaddress,twocolumn]{revtex4}
\usepackage{graphicx}
\usepackage{amsfonts}
\usepackage{url}
\usepackage{color}
 \usepackage{ulem}

\begin{document}

\title{Temperature and friction fluctuations inside a harmonic potential}              
\author{Yann~Lanoisel\'ee}
 \email{y.lanoiselee@bham.ac.uk}
\affiliation{
Institute of Metabolism and Systems Research, University of Birmingham, Birmingham B15 2TT, UK}
\affiliation{
Centre of Membrane Proteins and Receptors (COMPARE), Universities of Nottingham and Birmingham, Birmingham B15 2TT, UK}

\author{Aleksander~Stanislavsky}
 \email{a.a.stanislavsky@rian.kharkov.ua}
\affiliation{
Institute of Radio Astronomy, 4 Mystetstv St., 61002 Kharkiv, Ukraine }

\author{Davide~Calebiro}
 \email{d.calebiro@bham.ac.uk}
\affiliation{
Institute of Metabolism and Systems Research, University of Birmingham, Birmingham B15 2TT, UK}
\affiliation{
Centre of Membrane Proteins and Receptors (COMPARE), Universities of Nottingham and Birmingham, Birmingham B15 2TT, UK}

\author{Aleksander~Weron}
 \email{aleksander.weron@pwr.edu.pl}
\affiliation{Faculty of Pure and Applied Mathematics, Hugo Steinhaus Center, Wrocław University of Science and Technology, Wyb. Wyspiańskiego 27, 50-370 Wroc\'{l}aw, Poland }

\date{\today}

\begin{abstract}

In this article we study the trapped motion of a molecule undergoing diffusivity fluctuations inside a harmonic potential. For the same diffusing-diffusivity process, we investigate two possible interpretations. Depending on whether diffusivity fluctuations are interpreted as temperature or friction fluctuations, we show that they display drastically different statistical properties inside the harmonic potential.
We compute the characteristic function of the process under both types of interpretations and analyse their limit behavior.
Based on the integral representations of the processes we compute the mean-squared displacement and the normalized excess kurtosis.
In the long-time limit, we show for friction fluctuations that the probability density function (PDF) always converges to a Gaussian whereas in the case of temperature fluctuations the stationary PDF can display either Gaussian distribution or generalized Laplace (Bessel) distribution depending on the ratio between diffusivity and positional correlation times.
\end{abstract}





\maketitle

\section{Introduction}
The description of molecular diffusion in heterogeneous media is a long-standing collective endeavour. 
With the development of advanced microscopy techniques \cite{Manzo2015,Shashkova2017,Lelek2021} and single-particle tracking algorithms \cite{Jaqaman2008,Mortensen2010,Tinevez2017,Speiser2021}, it is now possible to record the diffusive motion of individual molecules with high spatial and temporal resolution. A number of methods have been developed to analyze such scenarios, see \cite{Vestergaard2014,Burnecki2014,Hoze2015,Janczura2020,MunozGil2021,Verdier2021,Lanoiselee2021,J.Szwabinski2022} and references therein. 
The progress in experimental diffusion measurements has fostered physical modelling of observed phenomena such as anomalous diffusion \cite{Weigel2010,J.Klafter2012}, which in turn has allowed a more quantitative description of biological phenomena \cite{Hoze2012,Sungkaworn2017,Weron2019,Calebiro2021}. Cases of transient anomalous diffusion have been observed \cite{Bronstein2009} and studied as well \cite{Saxton2007}.

Recently, a new phenomenon under the name `anomalous yet Brownian' diffusion has been discovered, whereby the displacement probability density function (PDF) of diffusive particles in a complex medium displays exponential tails as opposed to the usual Gaussian distribution. In most cases, the PDF shows exponential tails at short times and converges to a Gaussian PDF at long times. It also displays large fluctuations of the time-averaged mean squared displacement \cite{Uneyama2015,Grebenkov2019}. Most notably, Granick and co-workers \cite{Wang2009,Wang2012} were the first to discover such anomalous yet Brownian diffusion phenomenon.
Next, Chubinsky and Slater \cite{Chubynsky2014} introduced the now popular diffusing diffusivity model, in which the diffusion coefficient of the tracer particle evolves in time like the position of a Brownian particle in a potential field. Then, Jain and Sebastian  formalized the diffusing diffusivity model using a path integral approach, which they explicitly solved in two spatial dimensions \cite{Jain2016}. This model has been further studied by Chechkin and co-workers see \cite{Chechkin2017} using the subordination technique. The model used in the present paper and introduced in \cite{Lanoiselee2018a} is a natural generalisation of the previous model \cite{Jain2016,Chechkin2017}. However, the dynamical foundations of nonextensive statistical mechanics were analysed much earlier in \cite{Beck2001}. Chechkin et al. also described a general method to build diffusing diffusivity from a Gaussian process in \cite{Sposini2018a} and applied it to fractional Brownian motion \cite{Wang2020}. The question of fluctuating diffusivity has also been studied by Miyaguchi and Akimoto \cite{Uneyama2015,Miyaguchi2016,Miyaguchi2019} who applied it to two-state diffusivity models as well as diffusing diffusivity. 
 One can also bridge the gap between multi-state diffusivity and diffusing diffusivity with the choice of a suitable state transition matrix \cite{Grebenkov2019}. The simplest model of integrated diffusing diffusivity (without memory), the `continuous-time random integrated diffusivity', was shown to display exponential tails on the extremities of the distribution at all times which become virtually invisible at long time such that the PDF converges to Gaussian distribution, as long as diffusivity increments have finite moments and exponential tails \cite{Lanoiselee2019}. Furthermore, it has been shown by Barkai and Burov \cite{Barkai2020} that exponential tails exhibit a universal behavior
based on a large deviation approach to continuous-time random walk. Cases of space dependent diffusivity have also been studied \cite{Luo2018,Postnikov2020}.

The Stokes-Einstein relationship expresses the diffusion coefficient $D$ as a function of other physical quantities, namely
\begin{equation}\label{eq:D_definition}
D=\frac{k_BT}{\gamma}\,,
\end{equation}
where $k_B$ is the Boltzmann constant, $T$ the temperature, and $\gamma$ denotes the friction coefficient. The expression for the friction coefficient takes the form $\gamma=6\pi\eta r$, where $\eta$ is the medium viscosity, and $r$ is the hydrodynamic radius the particle.
There is a wealth of studies that consider the impact of fluctuating diffusivity on the statistical properties of a particle diffusing without the presence of a force. Without external force, fluctuations of either of these values are indistinguishable. However, in the presence of a potential, the potential-derived force is scaled by friction and does not depend on temperature.
In general, the Langevin equation with an external potential is written as
\begin{equation}\label{eq:Lagevin_diffusion_potential}
    dx_t=-\frac{1}{\gamma}\Delta(V(x_t))dt+\sqrt{2D}\,dW_t\,,
\end{equation}
where $\Delta(V(x_t)$ is the gradient of the potential at position $x_t$.
Here, the dynamics will be affected differently depending on whether it is temperature or friction that do fluctuate.
This gives an opportunity to tell friction from temperature fluctuations apart.\\
In the context of diffusion in living cells, trapping of laterally diffusing molecules on the plasma membrane is relevant to signal transduction. Receptors at the plasma membrane mediate intracellular downstream signalling pathways upon their stimulation with the proper external stimulus. 
It has been shown that receptor\textendash effector interaction is increased in the presence of nanodomains at plasma membrane, where both molecule types are confined \cite{Sungkaworn2017}.
Depending on the receptor, there are different candidates for the nature of these domains, whether they are phase-separated lipid domains \cite{Cebecauer2018}, or being defined by structural components like clathrin-coated pits \cite{Cocucci2012} or actin delimited barriers or even anchor points on actin filaments \cite{Kusumi2005}.\\
Physically, many mechanisms have been invoked to explain confinement. First, molecules can be enclosed in a boundary-delimited space \cite{Taflia2007}. An example of this is the actin network underlying the cell plasma membrane, which can act as a barrier for membrane proteins \cite{Kusumi2005}. These phenomena can be described as diffusion inside a domain with reflecting boundary condition \cite{Lanoiselee2018b,Sposini2018,Lanoiselee2019,Grebenkov2021}. In this case, there is no force exerted on receptors inside the domain such that the temperature fluctuations are also indistinguishable from friction ones.
Another source of trapping can be the presence of a potential well that attracts surrounding molecules. The attracting potential can be due to the presence of a specific molecule or to a particular composition of the local environment. The simplest physical model for this trapping is the harmonic potential defined by 
\begin{equation}
  V(x)=\frac{k}{2}(x-\bar{x})^2\,,
\end{equation}
 where $k$ is the spring constant. When the diffusion coefficient is constant, this case is known as the Ornstein-Uhlenbeck (OU) process. 
In living cells, molecules are compartmentalized into nano-domains. In these nano-domains many factors can affect the diffusion coefficient.
The hydrodynamic radius $r$ of the molecule can fluctuate \cite{Yamamoto2021} due to conformation changes. The temperature $T$ can vary locally due to either endothermic or exothermic chemical reactions in the vicinity \cite{Okabe2018,Oyama2020,Balaban2020}, when the viscosity can be affected by the bulk composition.
Confinement of diffusive molecules being a key feature in living cells, we aim to go one-step further in its statistical description. The effect of diffusivity fluctuations inside a harmonic potential remains poorly understood apart from the study \cite{Uneyama2019} on relaxation functions in the case of friction fluctuations. We wish to investigate the effects of local fluctuations of either temperature or friction within the trapping domain.  In both cases, the fluctuations of these quantities can be expressed in terms of a fluctuating diffusivity around its equilibrium value. For both fluctuation types to be comparable, we impose that both models share the same diffusivity process yet with different interpretations. 

We show that two dimensionless parameters are sufficient to summarize the behavior in both cases. The first $\nu$ quantifies the strength of diffusivity fluctuations 
\begin{equation}
    \nu=\frac{\bar{D}}{\sigma^2\tau_D}\,.
\end{equation}
It compares the average diffusion coefficient $\bar{D}$ to the average amplitude of diffusivity fluctuations $\sigma^2\tau_D$, where $\sigma$ controls the amplitude of diffusivity stochastic component, and $\tau_D$ is the `diffusivity correlation time'. Large values $\nu\gg 1$ denote almost constant diffusivity, whereas small values $\nu\ll 1$ manifest large fluctuations.
The second parameter $\mu$
 compares the `positional correlation time' $\tau_x$  to the diffusivity correlation time $\tau_D$ as
\begin{equation}
    \mu=\frac{\tau_x}{2\tau_D}\,.
\end{equation}
This quantity shows whether diffusivity ($\mu>1/2$) or position ($\mu<1/2$) equilibrium is reached faster.

First, in Sec. \ref{sec:temperature_fluctuations} we investigate the case of a molecule in a harmonic potential, where the fluctuating diffusion coefficient is interpreted as temperature fluctuations. We derive the exact characteristic function of the process and study the probability density function of displacements in the long-time limit. Then, we proceed with computing the mean squared displacement as well as the long-time behavior of normalized excess kurtosis of this process and demonstrate its weak ergodicity property. So far, it has been possible to obtain a stationary probability density function with exponential tails, but at the cost of adding discontinuity in the motion of the particle. The first example is stochastic resetting \cite{ems20,Stanislavsky2022}, where particle returns to the origin at random times. The second example is a model of subordinated random walks with the Laplace exponent being the conjugated inverse stable subordinator \cite{Stanislavsky2021} which is a pure jump process. Here, we will describe conditions under which temperature fluctuations leads to a stationary PDF with exponential tails while ensuring continuity of the displacement. 

Next, in Sec. \ref{sec:friction fluctuations} we investigate the case of diffusivity fluctuations interpreted as friction fluctuations. In this case, the process can be recast as a subordinated OU process. We study its second moment and normalized excess kurtosis. We show that, similarly to diffusing-diffusivity models without force, the stationary PDF is Gaussian in any case. However, while the second moment is unchanged without force, here all the moments are strongly affected by friction fluctuations. In this case, we also prove the weak ergodic behavior of the process.

 Finally, we highlight the results and verify analytical solutions with numerical simulations.

\section{Diffusive model in harmonic potential -- the case of temperature fluctuations}
\label{sec:temperature_fluctuations}
We present a model where molecules are trapped within a confining potential with a fluctuating time-dependent temperature $T_t$. In this case, $T_t$ is a function of diffusivity, i.\,e.
\begin{equation}
    T_t=\frac{D_t\gamma}{k_B}\,.
\end{equation}
The diffusion in a harmonic potential is modelled with an OU process with mean position $\bar{x}$ and correlation time $\tau_x$, where diffusivity is time-dependent. 
To model temperature fluctuations we use a diffusing diffusivity process known as a Cox-Ingersoll-Ross process or a square root process.

The first term of the Langevin equation for diffusivity is a harmonic potential that drives diffusivity toward its average $\bar{D}$ with a correlation time $\tau_D$. The second term describes the fluctuations of diffusivity with strength $\sigma$ proportional to the square root of diffusivity.
When $D$ gets close to $0$, the fluctuations becomes smaller, thus ensuring non-negativity of $D_t$.

The coupled Langevin equation for the position $x_t$ and the diffusivity $D_t$ reads
\begin{equation}\label{eq:coupled_Langevin_temperature_fluctuation}
\left\{
    \begin{array}{ll}
dx_t=-\frac{1}{\tau_{x}}(x_t-\bar{x})dt+\sqrt{2D_t}\,d W_t^{(1)} \\
dD_t=-\frac{1}{\tau_{D}}(D_t-\bar{D})dt+\sigma\sqrt{2D_t}\,dW_t^{(2)}
    \end{array}\,,
\right.
\end{equation}
where 
\begin{equation}\label{eq:tx_fct-gamma-k}
    \tau_{x}=\frac{\gamma}{k}\,,
\end{equation} 
is the position correlation time, $\bar{x}$ and
\begin{equation}\label{eq:mean_D} \bar{D}=\frac{k_BT}{\gamma}
\end{equation}
are respectively the average position and the average diffusivity, and $\sigma$ is the 'speed' of fluctuation of the diffusion coefficient (here $T$ and $\gamma$ without subscript $t$ denote the average values). Note that the two Wiener processes are independent, i.\,e. $\langle dW_t^{(1)}dW_t^{(2)}\rangle=0$.

One can show that for $\nu\geq 1$ we have $D_t>0$ while in the case of $\nu<1$ the diffusivity may reach $D_t=0$. To ensure strict positivity as required for any diffusion coefficient to have physical meaning, we introduce a reflecting boundary condition at $D=0$.
This diffusing diffusivity model \cite{Lanoiselee2018a,Lanoiselee2018b} is a generalisation of the model based on the squared distance from the origin of a $n$-dimensional OU process from \cite{Jain2016,Chechkin2017,Tyagi2017,Jain2017}, in which the value $\nu$ was limited to integer values only.

We emphasize that in the case of temperature fluctuations, Eq. \ref{eq:Lagevin_diffusion_potential} cannot be reduced to a subordination scheme of the OU process as studied in the case of the inverse stable subordinator \cite{Gajda2015}. However, this approach can be used in the case of friction fluctuations.

The corresponding forward Fokker-Planck equation for the joint probability $P(x,D,t|x_0,D_0)$ of being at position $x$ and diffusivity $D$ at time $t$ and starting from $x_0,D_0$ has the following form
\begin{eqnarray}\nonumber
\frac{\partial P(x,D,t\vert x_0,D_0)}{\partial t}&=&
D\frac{\partial^2}{\partial x^2}P+\sigma^2\frac{\partial^2}{\partial D^2}\left(DP\right)
\\\nonumber
&+&\frac{1}{\tau_{x}}\frac{\partial}{\partial x}\left[(x-\bar{x})P\right]\nonumber\\
&+&\frac{1}{\tau_{D}}\frac{\partial}{\partial D}\left[(D-\bar{D})P\right],
\end{eqnarray}
with the initial condition $P(x,D,0\vert x_0, D_0)=\delta(x-x_0)\delta(D-D_0)$.
We perform the Fourier transform for the coordinate $x$ and the Laplace transform with respect to the variable $D$  through the general integral transform
\begin{eqnarray}
    P^*(q,s,t|x_0,D_0)&=&\int\displaylimits_{-\infty}^\infty dx \int\displaylimits_{0}^\infty dD\,e^{-sD-iqx}\nonumber\\ 
    &\times&P(x,D,t|x_0,D_0)\,.
\end{eqnarray}
The detailed derivation of the characteristic function can be found in Appendix \ref{section: Full derivation}. Being unable to measure directly the value $D_t$ over time in a real experiment, we average over $D_0$ and $D$. Then, we deduce the characteristic function $P^*(q,t|x_0)$ associated with the marginal probability density $P(x,t|x_0)$ that gives
\begin{eqnarray}\label{Eq:char_fun_marginal}
P^*(q,t|x_0)&=&
    \exp\left(-iq\left(\bar{x}+
    (x_0-\bar{x})e^{-t/\tau_x}\right)\right)\nonumber\\
   &\times&\left(\frac{e^{t/\tau_D}/b}{F_1(b,t)
   -
   \frac{\bar{D}}{\sigma}|q|
   e^{-t/\tau_x}F_2(b,t)}\right)^\nu,
\end{eqnarray}
with 
\begin{eqnarray}
F_1(b,t &=& I_{-\mu}(be^{-t/\tau_x})K_{1-\mu}(b)\nonumber\\
        &+& K_{\mu}(be^{-t/\tau_x})I_{1-\mu}(b),
\end{eqnarray}
and
\begin{eqnarray}
F_2(b,t) &=&I_{1-\mu}(be^{-t/\tau_x})K_{1-\mu}(b)\nonumber\\
        &+&K_{1-\mu}(be^{-t/\tau_x})I_{1-\mu}(b),   
\end{eqnarray}
where $b=\sigma\tau_x|q|$, where the value $\sigma\tau_x$ plays the role of a length-scale. Here $K_\alpha(z)$ and $I_\alpha(z)$ are the modified Bessel functions of the second kind \cite{abr64}.
Whereas the first exponential term in Eq.(\ref{Eq:char_fun_marginal}) corresponds to the average position, the second factor encompasses the intricate dynamics of temperature fluctuations with the mean-reverting behavior of the positional OU component.

\subsection{Long-time behavior and its limiting forms}
While the characteristic function in Eq.(\ref{Eq:char_fun_marginal}) is exact and valid at all times, its behavior is not easy to grasp. In order to better understand the PDF corresponding to Eq.(\ref{Eq:char_fun_marginal}), we focus on the long-time limit when the system reaches equilibrium.
To do so, we first compute the characteristic function $P^*(q,t|x_0)$ in the long-time limit $t\to\infty$ and then study its limiting behavior.
We use the small $z$ argument expansion of $K_\alpha(z)$ and $I_\alpha(z)$ to obtain the long-time characteristic function
  \begin{flalign}\label{Eq:char_fun_long_time}
  P^*(q)=
      \exp\left(-iq\bar{x}\right) 
  \left(
  \frac{
         (b/2)^{\mu-1}
        }{\Gamma(\mu)
        I_{\mu-1}(b)
        }
  \right)^{\nu}.
  \end{flalign}
  This expression is much simpler than Eq.(\ref{Eq:char_fun_marginal}).
  To develop a better understanding of this characteristic function, we consider the limit behavior when the correlation time of diffusivity is much larger than the correlation time of position ($\mu\ll 1$) as well as the reverse case ($\mu\gg 1$).\\
  First, we reason that when $\mu\ll 1$, the correlation time of diffusivity is much longer than the correlation time of position, such that the diffusivity remains nearly constant for a particle while reaching positional equilibrium. Therefore, particles with small $D$ will be less able to fight the attracting force in comparison to molecules with a large $D$. As a result, for each value of $D$ there is a different conditional stationary PDF $P_\infty(x|D)$. For a specific $D$ our model is simply an OU process with the diffusion coefficient $D$. Therefore, we use the stationary regime of the OU process for the conditional PDF $P_\infty(x|D)$ written as
  \begin{equation}\label{Eq:conditional_prob_small_mu}
      P_\infty(x|D)=\frac{1}{\sqrt{2\pi D \tau_x}}\exp\left(-\frac{(x-\bar{x})^2}{2D\tau_x}\right)\,.
  \end{equation}
 Figure (\ref{fig:sim_inf_fct_tau_x_over_tau_D}.A) shows the perfect agreement between the conditional probability Eq. (\ref{Eq:conditional_prob_small_mu}) and simulation.
 To get the marginal probability $P_\infty(x)$ we average over $D$
 \begin{equation}
     P_\infty(x)=\int\displaylimits_0^\infty P_\infty(x|D)p_\infty(D)\,,
 \end{equation}
 where $p_\infty(D)$ - the stationary PDF of $D$ - corresponds to
  \begin{equation}
      p_\infty(D)=\frac{\nu^\nu}{\bar{D}^\nu\Gamma(\nu)}D^{\nu-1}\exp \left(-\frac{\nu}{\bar{D}}D \right)\,.
  \end{equation}
Averaging over $D$ yields
\begin{eqnarray}\label{Eq:PDF_small_mu}
      P_\infty(x)&=&\frac{2^{-\nu/2+3/4}\nu^{1/2}}{\sqrt{\pi\bar{D}\tau_x}\Gamma(\nu)}\left(\sqrt{\frac{\nu}{\bar{D}\tau_x}}|x-\bar{x}|\right)^{\nu-1/2}\\\nonumber
&\times&      K_{\nu-1/2}\left(\sqrt{\frac{2\nu}{\bar{D}\tau_x}}|x-\bar{x}|\right),
  \end{eqnarray}
with $P_\infty(0)=\frac{\Gamma(\nu-1/2)}{\Gamma(\nu)}\sqrt{\frac{\nu}{2\pi\bar{D}\tau_x}}$. This distribution is known as the generalised Laplace or variance gamma \cite{Madan1990}, or K-distribution. It is useful for modelling share price returns, where existing choices have shortcomings \cite{Madan1998}. Often, the price data show that returns of financial assets are actually skewed and have higher kurtosis than would be expected. This means that the data is heavier in tails and have a higher centre, more ``peaked'' than a normal distribution. 
In particular, for the case $\nu=1$, we have the Laplace distribution
    \begin{equation}
      P_\infty(x)=\frac{1}{\sqrt{2\bar{D}\tau_x}}
e^{-\sqrt{\frac{2}{\bar{D}\tau_x}}|x|}\, ,
  \end{equation}
  for more information on the Laplace distribution see Appendix \ref{section: Laplace distribution}.
Interestingly, at small space frequencies $q\to 0$, the characteristic function in Eq.(\ref{Eq:char_fun_long_time}) yields the expression
    \begin{flalign}\label{Eq:char_fun_small_mu}
  P^*(q)=
  \frac{ \exp\left(-iq\bar{x}\right)  }{
      \left(1+\eta^2q^2\right)^\nu
        }\,,
  \end{flalign}
 where $\eta^2=\bar{D}\tau_x/(2\nu)$, which corresponds exactly to the characteristic function in the case $\mu\to 0$ and to the PDF in Eq.(\ref{Eq:PDF_small_mu}) as illustrated in Fig.(\ref{fig:sim_inf_fct_tau_x_over_tau_D}.B). This proves the non-Gaussian character of the distribution for small $\mu$ and finite $\nu$. Indeed, in the limit $\nu\to\infty$, the characteristic function Eq.(\ref{Eq:char_fun_small_mu}) becomes that of the OU process with the diffusion coefficient $\bar{D}$, satisfying
    \begin{flalign}\label{eq:temp_small_mu_large_nu}
  P^*(q)\sim e^{-iq\bar{x}} 
       e^{-q^2\bar{D}\tau_x/2}\,.
  \end{flalign}
To study the case when $\mu\gg 1$, in Appendix \ref{appendix:BesselI_beta_large_order} we compute the large order expansion of $I_\beta(b)$ from which, after simplification, we get
the same PDF as in Eq.(\ref{eq:temp_small_mu_large_nu}).
We conclude that in the limit $\mu\to\infty$, the distribution is Gaussian and centered on $\bar{x}$  with constant diffusion coefficient $\bar{D}$, the same stationary distribution as the usual OU process.

Our interpretation of this result is that particles explore the possible diffusivities faster than the time needed to reach positional equilibrium within the harmonic potential. Therefore, diffusivity is averaged out at equilibrium such that the position is independent of $D$ and depends only on the average diffusivity $\bar{D}$. This is illustrated in Fig.(\ref{fig:sim_inf_fct_tau_x_over_tau_D}.C), where the conditional PDF $P(x|D)$ of simulated data is in perfect agreement for any value of $D$ with Eq.(\ref{eq:temp_small_mu_large_nu}).

From the presented results, we conclude that the model is very general and can therefore accommodate for a large variety of PDF shapes. This is illustrated in Fig.(\ref{fig:sim_inf_fct_tau_x_over_tau_D}.D) where the stationary PDF for $\nu=0.1$ and different values of $\mu$.

\subsection{Short-time behavior}
Next, we compare the PDF shapes for the initial and the stationary conditions.
Starting from the center of the well $x_0=\bar{x}$, at short times $t\ll \tau_x$ the diffusion is unaffected by the potential so that the
position can be approximated by the ordinary Brownian motion with the initial diffusivity $D_0$, having
\begin{equation}
x_t=\int\displaylimits_0^t\sqrt{2D_0}dW_s\,.
\end{equation}
Then the marginal PDF $P_0(x,t)$ reads
\begin{equation}
    P_0(x,t)=\int\displaylimits_0^\infty
    P_0(x,D_0,t)p_\infty(D_0)dD_0'\,,
    \end{equation}
which can be found exactly
  \begin{eqnarray}
      P_0(x)&=&\frac{2^{1/2-\nu}\nu^{1/2}}{\sqrt{\pi\bar{D}t}\Gamma(\nu)}\left(|x|\sqrt{\frac{\nu}{\bar{D}t}}\right)^{\nu-1/2}\nonumber\\
      &\times&K_{\nu-1/2}\left(|x|\sqrt{\frac{\nu}{\bar{D}t}}\right)\,.
  \end{eqnarray}
  
In the case of $\mu\ll 1$ the distribution conserves the same shape over time, even if the length-scale of the process is changed. In turn for $\mu\gg 1$, the initial shape of the distribution is completely lost when positional equilibrium has occurred. Note that in the case of friction fluctuations (Sec. \ref{sec:friction fluctuations}), the initial distribution is exactly the same, however we will show that the long time behavior is different  
\subsection{It$\hat{\rm o}$ calculus and moments}
In this section, we use the integral representation of the processes to compute moments and the normalized excess kurtosis. The integral representation for the position of a particle is similar to that of an OU process yet with time-dependent diffusivity
\begin{eqnarray}
 x_t&=&\bar{x}+ (x_0-\bar{x})e^{-t/\tau_x}\nonumber\\
 &+&e^{-t/\tau_x}\int\displaylimits_0^te^{s/\tau_x}\sqrt{2D_s}dW_s^{(1)}\,.
\end{eqnarray}
The properties of the diffusivity process (integral representation, mean, second moment, autocorrelation) have already been studied in \cite{Lanoiselee2018a}.

The mean position is not affected by temperature fluctuations and reads
\begin{equation}
    \langle x_t \rangle=\bar{x}+ (x_0-\bar{x})e^{-t/\tau_x}\,.
\end{equation}
The second moment is equal to
\begin{eqnarray}\label{Eq:second_moment_fct_D_0}
    \langle x_t^2\rangle
    &=&\bar{D}\tau_x\left(1-e^{-2t/\tau_x}\right)
    +\langle x_0^2\rangle e^{-2t/\tau_x}\\\nonumber
    &+&\frac{(\langle D_0\rangle-\bar{D})\tau_x}{(1-\mu)}\left(e^{-t/\tau_D}-e^{-2t/\tau_x}\right)\,,
\end{eqnarray}
where the two first terms correspond to the unperturbed OU process, while the third term is due to fluctuations of diffusivity.
Taking the diffusivity equilibrium, we have
\begin{equation}\label{Eq:second_moment_average}
    \langle x_t^2\rangle
    =\bar{D}\tau_x\left(1-e^{-2t/\tau_x}\right)\,.
\end{equation}




So far, the moments are very similar to that of the OU process.
The next step is to go beyond the second moment to deduce in which regime the PDF is Gaussian-like or not.

The case of the fourth moment is more involved due to the complex intricacy of diffusivity fluctuations and the attractive force. Readers can refer to Appendix \ref{Appendix:fourth_moment} for a more detailed derivation. In the long-time limit, the fourth moment is
     \begin{equation}\label{eq:fourth_mom_long_time}
         \langle x^4(t\to\infty)\rangle=3\bar{D}^2\tau_x^2\left(1+\frac{1}{\nu(1+\mu)}\right)\,.
     \end{equation}
From the second and the fourth moments we deduce the normalised excess kurtosis $\kappa$ in the form
\begin{equation}
    \kappa(t)=\frac{1}{3}\frac{\langle X^4(t)\rangle}{\langle X^2(t)\rangle^2}-1\,,
\end{equation} 
which is equal to $1$ for the Laplace distribution and equal to $0$ for the Gaussian distribution. We combine Eq.(\ref{Eq:second_moment_fct_D_0}) and Eq.(\ref{eq:fourth_mom_long_time}) to obtain the long-time normalized excess kurtosis
     \begin{equation}
         \kappa(t\to\infty)=\frac{1}{\nu(1+\mu)}\,.
     \end{equation}
Both in the case of $\mu\to \infty$ - when diffusivity is averaged before reaching positional equilibrium - and in the case of $\nu\to\infty$ - when diffusivity is constant - the normalized excess kurtosis vanishes such that the distribution is Gaussian.
     
In turn, for $\mu \to 0$ corresponding to the PDF in Eq.(\ref{Eq:PDF_small_mu}), the normalized excess kurtosis equals $1/\nu$, and its shape is entirely governed by the amplitude of diffusivity fluctuations. 

\begin{figure}[h!]\label{fig:sim_inf_fct_tau_x_over_tau_D}
\includegraphics[width=\linewidth]{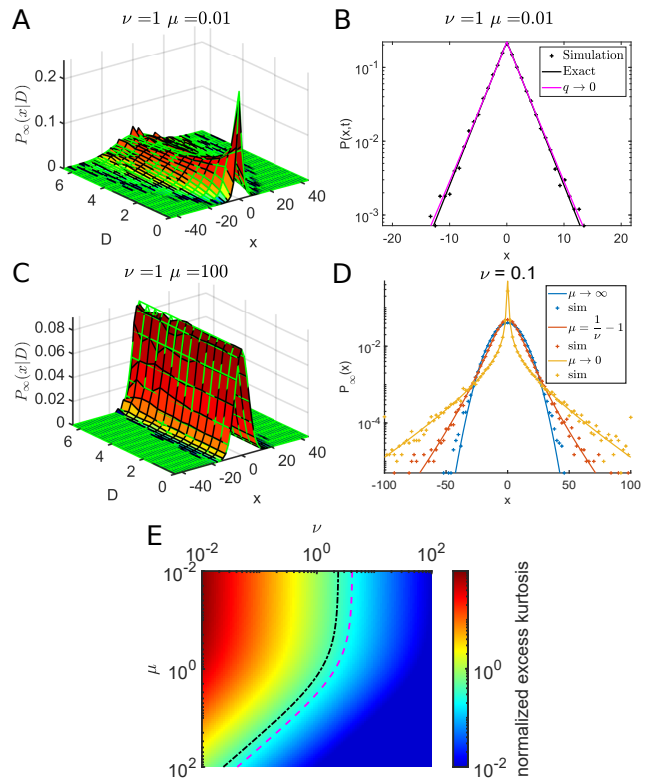}
\caption{
{\bf (A)}
Stationary conditional PDF $P_\infty(x|D)$ in the case of temperature fluctuations with parameters $\mu=0.1$, $\nu=1$, $\tau_x=20$ and $\bar{D}=1$. 
{\bf (B)}
Stationary PDF in the case $\mu=0.01$ based on simulation (black dots), overlayed with the exact expression (black curve) and with the small frequency behavior ($q\to 0$) corresponding to the limit $\mu\to 0$.
{\bf (C)}
Stationary conditional PDF $P_\infty(x|D)$ in the case of temperature fluctuations with parameters $\mu=10$, $\nu=1$, $\tau_x=20$ and $\bar{D}=1$. 
{\bf (D)}
Simulated (dots) and theoretical (lines) stationary PDF as a function of the ratio of correlation times taking values $\mu=0.01$ (yellow), $\mu=1/\nu-1=1$ (red), $mu=100$ (blue) with parameters $\nu=0.1$,$\tau_x=20$ and $\bar{D}=1$.
{\bf (E)} Normalized excess kurtosis of the long-time stationary PDF as a function of $\mu$ and $\nu$. Black dashed line corresponds to the value for which transition from Laplace to Gaussian is detected using \cite{kundu04} and purple dashed line corresponding to the value for which Jarque-Bera test \cite{Jarque1980} detects Gaussian PDF.
}
\end{figure}

Figure (\ref{fig:sim_inf_fct_tau_x_over_tau_D}.E) illustrates the theoretical normalized excess kurtosis in the stationary regime $\kappa(t\to\infty)$ as a function of $\mu$ and $\nu$. Additionally, the PDF was simulated with thousand of points and its Gaussianity was tested with two methods. The first method compares whether Laplace or Gaussian PDF explains better the data \cite{kundu04} while the second is the Jarque-Bera goodness-of-fit test that is based on kurtosis statistics.
Lines are drawn at the critical value of $\kappa(t\to\infty)$ from which both methods found a Gaussian distribution. 






\subsection{Ergodicity}
When temperature fluctuates, the system is generally out of equilibrium. However, in our case, temperature fluctuates around an average with a stationary distribution at long-time. Therefore, one can wonder whether this model shows ergodicity breaking or not.
In the case of infinitely divisible processes one can use the Wiener-Khintchine theorem to prove ergodicity if the autocorrelation function vanishes \cite{Khinchin1949,Lapas2008,Burov2010}. Other approaches make use of the dynamical functional \cite{{Magdziarz2011,Janczura2015,Lanoiselee2016}}. But in our case the process is not infinitely divisible so these tools are not suitable. We then question ergodicity in a weaker sense by determining the time-averaged mean square displacement and comparing it to the generalized MSD following the strategy developed in \cite{Mardoukhi2020} for the case of an OU process with constant diffusion coefficient.
For this we first compute the generalized MSD $\langle  (x_{t+\Delta}-x_t)^2 \rangle$.
Next, we compute the average over $x_0$ for which the process is assumed to start at equilibrium yielding $\langle x_0\rangle=\bar{x}$, $\langle x_0^2\rangle=\bar{D}\tau_x$ and $\langle D_0 \rangle=\bar{D}$. Thus, we have
\begin{equation}
    \langle  (x_{t+\Delta}-x_t)^2 \rangle =2\bar{D}\tau_x\left(1-e^{-\Delta/\tau_x}\right)\,.
\end{equation}
The ensemble averaged TAMSD follows
\begin{eqnarray}
    \langle \delta^2(\Delta,t)\rangle
    &=&
2\bar{D}\tau_x\left(1-e^{-\Delta/\tau_x}\right).
\end{eqnarray}

Finally, we compute the ergodicity breaking parameter \cite{He2008,Schwarzl2017}
\begin{equation}
EB(\Delta)= \lim_{t\to\infty}   \frac{\langle \delta^2(\Delta,t)\rangle}{    \langle (x_{t+\Delta}-x_t)^2 \rangle }-1=0
\end{equation}
which is equal to $0$, thus proving ergodicity in the weak sense. Despite temperature fluctuations, the second moment is ergodic.


\section{Diffusive model in harmonic potential -- The case of fluctuating friction coefficient}\label{sec:friction fluctuations}
In this section we study the case of a particle diffusing in a harmonic potential, where friction $\gamma_t$ fluctuates over time while temperature remains constant. 
To apply the same model for diffusivity as in Sec. \ref{sec:temperature_fluctuations}, we use the relationship described in Eq. (\ref{eq:D_definition}) to express the time-dependent friction coefficient $\gamma_t$ as a function of $D_t$ in the form 
\begin{equation}
\gamma_t=\frac{k_B T}{D_t}\,.
\end{equation}
In this case, the coupled Langevin equation for position $x_t$ and diffusivity $D_t$ is written as
\begin{equation}
\left\{
    \begin{array}{ll}
dx_t=-\frac{1}{\tau_x}\frac{D_t}{\bar{D}}(x_t-\bar{x})dt+\sqrt{2D_t}\,d W_t^{(1)} \\
dD_t=-\frac{1}{\tau_D}(D_t-\bar{D})dt+\sigma\sqrt{2D_t}\,dW_t^{(2)}
    \end{array}\,,
\right.
\end{equation}
where the inverse positional correlation time $\frac{1}{\tau_x}\frac{D_t}{\bar{D}}$ is obtained by combining Eq.(\ref{eq:tx_fct-gamma-k}) and Eq.(\ref{eq:mean_D}). It is clear that, contrarily to Sec. \ref{sec:temperature_fluctuations}, the inverse correlation time fluctuates around its mean $1/\tau_x$ in the same way as diffusivity does fluctuate around $\bar{D}$. Note that the equation for diffusivity is identical to  Eq.(\ref{eq:coupled_Langevin_temperature_fluctuation}). 

To study this equation, we
rescale time by diffusivity $dt^*=D_tdt$, where $t^*$ has a unit of integrated diffusivity ($m^2$). This variable change allows one to treat the process within the subordination framework \cite{Chechkin2017}.
The rescaled equation gives
\begin{equation}
\left\{
    \begin{array}{ll}
dx_{t^*}=-\frac{1}{\bar{D}\tau_x}(x_{t^*}-\bar{x})dt^*+\sqrt{2}\,d W_{t^*}^{(1)} \\
dt^*=D_t dt,\\
dD_t=-\frac{1}{\tau_{D}}(D_t-\bar{D})dt+\sigma\sqrt{2D_t}\,dW_t^{(2)}
    \end{array}\,,
\right.
\end{equation}
where the first equation corresponds to the parent process that is an OU process and the second equation corresponds to the subordinator that defines the integrated diffusivity $t^*=\int_0^tD_sds$.

The probability density function of the parent process $p(x,t^*)$ for the position $x$ as a function of the subordinator $t^*$
is Gaussian with mean $\langle x_{t^*}\rangle=\bar{x}+(x_0-\bar{x})e^{-t^*/(\bar{D}\tau_x)}$
and variance $\langle (x_{t^*}-\langle x_{t^*}\rangle)^2\rangle= \bar{D}\tau_x\left(1-e^{-2t^*/(\bar{D}\tau_x)}\right)$.
The corresponding characteristic function of the parent process takes the form
\begin{eqnarray}
    \tilde{p}(q,t^*)&=&e^{-iq\left[\bar{x}+(\bar{x}-x_0)e^{-t^*/(\bar{D}\tau_x)}\right]}\nonumber\\
    &\times&e^{-\frac{q^2}{2}\bar{D}\tau_x(1-e^{-2t^*/(\bar{D}\tau_x)})}\,.
\end{eqnarray}

To obtain the characteristic function of the process, one needs to integrate the characteristic function $\tilde{p}(q,t^*)$ of the parent process over the probability density $\Pi(t^*,t)$ of integrated diffusivity, namely
\begin{equation}\label{eq:char_fun_friction_integral_rep}
    \tilde{P}(q,t)=\int_0^\infty \tilde{p}(q,t^*)\Pi(t^*,t)dt^*.
\end{equation}
Unfortunately, the exact expression for $\Pi(t^*,t)$ is unknown, as the integral cannot be computed explicitly. However, the integral representation of the characteristic function in Eq.(\ref{eq:char_fun_friction_integral_rep}) will be useful to find the moments of the process in the next section.

\subsection{Moments and normalized excess kurtosis}
Now we study the effect of friction fluctuations on the moments and the normalized excess kurtosis of the process. At small values $q$ the characteristic function reads
\begin{eqnarray}
    &&\tilde{P}(q,t)
    \sim 1
    -iq\int_0^\infty \left[\bar{x}+(\bar{x}-x_0)e^{-\frac{t^*}{(\bar{D}\tau_x)}}\right]\Pi(t^*,t)dt^*
    \nonumber\\
    &&-q^2\frac{\bar{D}\tau_x}{2}\int_0^\infty\left(1-e^{-2t^*/(\bar{D}\tau_x)}\right)\Pi(t^*,t)dt^*\\
    &&+
     \frac{q^4}{2}\left(\frac{\bar{D}\tau_x}{2}\right)^2\int_0^\infty\left(1-e^{-2t^*/(\bar{D}\tau_x)}\right)^2\Pi(t^*,t)dt^*\,.\nonumber
\end{eqnarray}

Using the formula for the moments $\langle x^k(t)\rangle=i^{-k}\frac{d^k\tilde{P}(q,t)}{dq^k}|_{q=0}$, we deduce the first moment
\begin{equation}
    \langle x_t \rangle=\bar{x}+(\bar{x}-x_0)
\hat{\Pi}(s,t)\Bigg|_{s=1/(\bar{D}\tau_x)},
\end{equation}
where $\hat{\Pi}(s,t)$ stand for the Laplace transform of the integrated diffusivity PDF.
Similarly, one can find the second moment, i.\,e.
\begin{equation}
\langle x_t^2\rangle=
\bar{D}\tau_x\left(1-\hat{\Pi}(s,t)\Big|_{s=2/(\bar{D}\tau_x)}\right).
\end{equation}
Both first and second moments are strongly affected by friction fluctuations (as opposed to the temperature fluctuation case) because the positional correlation time is fluctuating.

\begin{figure}[h!]\label{fig:sim_temp_versus_friction}
\includegraphics[width=\linewidth]{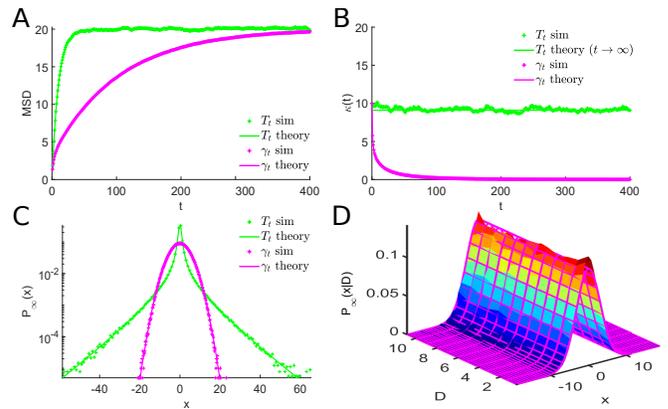}
\caption{
{All the results presented in this figure have been obtained with parameters $\bar{D}=1$, $\nu=0.1$, $\tau_x=20$ and $\mu=0.1$. \bf (A)}
Simulation (dots) overlayed with theory (lines) for the MSD in the case of temperature fluctuations (green) and friction fluctuations (magenta).
{\bf (B)}
Simulation overlayed with theoretical curve for the normalised excess kurtosis $\kappa(t)$ in the case of temperature fluctuations (green) and friction fluctuations (magenta). Parameters are $\nu=1$, $\mu=0.01$, $tau_x=20$ and $\bar{D}=1$.
{\bf (C)}
Simulation overlayed with theoretical curve for the the stationary PDF in the case of temperature fluctuations (green) and friction fluctuations (magenta).
{\bf (D)}
Simulation (colored surface) overlayed with theory (magenta mesh) for the stationary conditional probability $P_\infty(x|D)$ in the case of friction fluctuations.
}
\end{figure}

For our model, it is known 
 \cite{Dufresne1990,Lanoiselee2018a} that
\begin{eqnarray}
\hat{\Pi}(s,t|D_0)&=&\left[\frac{e^{\frac{t}{2\tau_D}}}{\cosh(\frac{\omega_s t}{2\tau_D})+\frac{1}{\omega_s}\sinh(\frac{\omega_s t}{2\tau_D})}\right]^{\nu}\\\nonumber
&&\times\exp\left[-\frac{sD_0\tau_D}{\omega_s}
\frac{2\sinh(\frac{\omega_s t}{2\tau_D})}
{\cosh(\frac{\omega_s t}{2\tau_D})+\frac{1}{\omega_s}\sinh(\frac{\omega_s t}{2\tau_D})}\right],
\end{eqnarray}
with $\omega_s=\sqrt{1+4s\sigma^2\tau_D^2}$.
Averaging over $D_0$, this expression yields

\begin{eqnarray}
\hat{\Pi}(s,t)&=&
\left(\frac{2e^{-\alpha_st}}{
1+e^{-\frac{\omega_s t}{\tau_D}}}\right)^\nu\nonumber\\
&\times&\frac{1}{\left(1
+\frac{1}{\omega_s}\left(1+s\frac{2\bar{D}\tau_D}{\nu}
\right)\tanh\left(\frac{\omega_s t}{2\tau_D}\right)
\right)^\nu}
\end{eqnarray}
with $\alpha_s=( \omega_s-1)/(2\tau_D)$, and $\displaystyle\lim_{x\to\infty}\tanh(x)=1$. Here, friction fluctuations strongly affect the relaxation time to positional equilibrium. For an arbitrary number $p$, we have $\omega_{2p/(\bar{D}\tau_x)}=\sqrt{1+4p/(\mu\nu)}$ which explicitly depends on the product $\nu\mu$. In the case where $\mu$ is small, we have $\alpha_{2p/(\bar{D}\tau_x}\approx 2\sqrt{p\nu/(\tau_D\tau_x)}$ such that the relaxation does not only depend on $\tau_x$ but also on the product with $\tau_D$ thus explaining the much slower relaxation of friction fluctuations compared to temperature fluctuations as illustrated in Fig.(\ref{fig:sim_temp_versus_friction}.A) using the same parameters for both cases.

However, when the lengthscale of thermal fluctuations $\sqrt{\bar{D}\tau_x}$ is larger than the lengthscale associated with diffusivity fluctuations $\sigma\tau_D$, then diffusive molecules have enough time to average out diffusivity fluctuations. Thus, the expression of the MSD turns to that of temperature fluctuation case Eq.(\ref{Eq:second_moment_average}), but the PDF is Gaussian so the dynamic is that of a simple OU
process.\\

In any case, at long times the MSD reads
\begin{equation}
\langle x^2(t\to\infty)\rangle=\bar{D}\tau_x\,.
\end{equation}
Similarly the fourth moment is equal to
\begin{eqnarray}
\langle x_t^4\rangle &=& 3\left(\bar{D}\tau_x\right)^2\Big(1-2\hat{\Pi}(s,t)\Big|_{s=2/(\bar{D}\tau_x)}\nonumber\\
&+&\hat{\Pi}(s,t)\Big|_{s=4/(\bar{D}\tau_x)}\Big)\,.
\end{eqnarray}
From which we deduce the normalized excess kurtosis, namely
\begin{equation}
\kappa(t)= 
\frac{\hat{\Pi}(s,t)\Big|_{s=4/(\bar{D}\tau_x)}-\left(\hat{\Pi}(s,t)\Big|_{s=2/(\bar{D}\tau_x)}\right)^2}
{\left(1-\hat{\Pi}(s,t)\Big|_{s=2/(\bar{D}\tau_x)}\right)^2}\,,
\end{equation}
which vanishes in the long-time limit
\begin{equation}
\kappa(t\to\infty)=0\,.
\end{equation}
Figure (\ref{fig:sim_temp_versus_friction}.B) shows the decay of the normalized excess kurtosis in the case $\mu\ll 1$ while that of temperature fluctuation remains constant.
Moreover, the $2n$-th moment can be computed
\begin{eqnarray}
    \langle x^{2n}(t)\rangle&=&(2n-1)!!\left(\bar{D}\tau_x\right)^n\Big(1\nonumber\\
    &+&\sum_{p=1}^n(-1)^p{n \choose p}\hat{\Pi}(s,t)\Big|_{s=2p/(\bar{D}\tau_x)}\Big)\,,
\end{eqnarray}
where $n!!=n\times (n-2)\times (n-4)\ldots$ is the double factorial. Given that $\displaystyle\lim_{t\to\infty}\hat{\Pi}(s,t)=0$, all the even moments converge to those of a Gaussian distribution in the long-time limit, meaning that the stationary PDF for the position is Gaussian in all scenarios a shown in Fig.(\ref{fig:sim_temp_versus_friction}.C) in striking contrast with temperature fluctuation case. Additionally the conditional probability distribution is independent of $D$ and depends solely on $\bar{D}$ as illustrated in Fig.(\ref{fig:sim_temp_versus_friction}.D).

\subsection{Ergodicity}
To investigate the ergodic properties, we start with the generalized second moment of the subordinator


\begin{equation}
   \left\langle\left( x_{t^*+\Delta^*}-x_{t^*}\right)^2\right\rangle
   =
   2\bar{D}\tau_x\left(1-e^{-\Delta^*/(\bar{D}\tau_x)}\right)\,.
\end{equation}
    
For the parent process the time averaged MSD is equal to the MSD
     \begin{eqnarray}
  \delta(\Delta^*,t^*) &=& \left\langle\left( x_{t+\Delta}-x_{t}\right)^2\right\rangle\nonumber\\
  &=&
   2\bar{D}\tau_x\left(1-e^{-\Delta^*/(\bar{D}\tau_x)}\right)\,.
     \end{eqnarray}
Then we average over the integrated diffusivity probability density for $t^*$ at time $t$ and for $\Delta^*$ at time $\Delta$ to get
  \begin{eqnarray}
   \delta^2(\Delta,t)  =
   2\bar{D}\tau_x\left(1-\hat{\Pi}(s,\Delta)\Bigg|_{2/(\bar{D}\tau_x)}\right).
     \end{eqnarray}
As a  result, the ergodicity breaking parameter is zero. We conclude that for friction fluctuations as well, the process is ergodic in the weak sense.

\section{Conclusion}
In this article, we have investigated the motion of a particle trapped inside a harmonic potential with diffusing diffusivity. Two cases were considered. The first where diffusing diffusivity is interpreted as temperature fluctuations. In the second case, it was interpreted as friction fluctuations. We showed that, in both cases, two essential quantities are useful to describe the system. The first value is $\nu$, which defines the inverse strength of diffusivity fluctuations, whereas the second is $\mu$, which quantifies the ratio between the position and the diffusivity correlation times. In both cases, when $\nu\to\infty$, diffusivity becomes a constant process, and the usual OU process is recovered at all times. When $\mu\gg 1$, but $\nu$ remains finite, in both cases the initial PDF of displacement shows exponential tails with a shape determined by $\nu$. The models converge to OU because diffusivity has time to self-average before particles can reach positional equilibrium.\\
However, in all the intermediate cases (finite $\nu$ and $\mu$), their behavior is drastically different. In the case of temperature fluctuations, the stationary long-time PDF displays exponential tails, and in the limit case $\mu\to 0$ the PDF conserves the same shape as in the initial condition. However, the first and second moments are the same as for the OU case, while the fourth moment differs.
In turn, for friction fluctuations the long-time stationary PDF is Gaussian in any case while the moments departs from that of an OU process because of the fluctuating positional correlation time.\\
The main results of the paper are:
\begin{itemize}
\item non-Gaussian PDF with continuous model and its ergodic properties;
\item generalized Laplace distribution with a confining potential for temperature fluctuations;
\item in both studied cases the diffusion coefficient has exactly the same distribution, however we show that
depending on whether it is a friction or a temperature fluctuation, the statistical properties of the process are very different.
\end{itemize}
We anticipate that these new results will be instrumental in understanding what happens to trapped molecules in an experimental setup, and offer far more greater detail than previous methods. Indeed one could test the presence of either temperature of friction fluctuations and use the statistical properties described here to quantify these distinct types of fluctuations.
On the theoretical side, our results raise questions about the relationship between the temperature fluctuation case studied here and stochastic resetting that can yield similar stationary PDF with exponential tails.


\begin{acknowledgments}
A.S. kindly acknowledges a support of the Polish National Agency for Academic Exchange (NAWA PPN/ULM/2019/1/00087/DEC/1) and A.W. a support of Beethoven Grant No. DFG-NCN 2016/23/G/ST1/04083. D.C acknowledges support by a Wellcome Trust Senior Research Fellowship (212313/Z/18/Z).
\end{acknowledgments}.

\section*{APPENDICES}
\appendix

\section{Laplace distribution}
\label{section: Laplace distribution}

The emergence of Laplace (or double exponential statistics) like the Gaussian, for various random
observables in nature, engineering, and finance, is widespread. Many examples range from
the first law of errors \cite{Laplace1986} to Laplace motion \cite{Kotz2001}. The Laplace distribution suggests a much better model to describe observations  than the Gaussian distribution with common variance, because each observer-instrument has its own variability, and all the participants of observations together result in large errors. Moreover, the explanation of anomalous diffusion tending to the confinement with the Laplace distribution is that diffusive motion, also accompanied by multiple trapping events with infinite mean sojourn time, makes impossible to leave such traps \cite{Stanislavsky2021a}. The Laplace distribution is also occurred as a steady state of Brownian motion under Poissonian resetting \cite{ems20}.

It was shown recently \cite{Stanislavsky2021a} that the Laplace confinement is present in confined random motions of both G proteins and receptors in living cells. It should be pointed out that the confined distribution form depends on the PDF of the parent process used for subordination. If we take Brownian motion, then the confined distribution has the Laplace form. This means that the presented mechanism can manifest itself as a source of the origin of jumps in heterogeneous systems. It is interesting that L\'evy motion as a parent process produces another confinement having the Linnik distribution.

\section{Full derivation for the fluctuating temperature case}
\label{section: Full derivation}

The corresponding forward Fokker-Planck equation for the joint probability $P(x,D,t|x_0,D_0)$ of being at position $x$ and diffusivity $D$ at time $t$ starting from values $x_0, D_0$ is
\begin{eqnarray}
\frac{\partial P(x,D,t\vert x_0,D_0)}{\partial t}&=&\frac{1}{\tau_{x}}\frac{\partial}{\partial x}\left[(x-\bar{x})P\right]\nonumber\\ 
&+&D\frac{\partial^2}{\partial x^2}P+\frac{1}{\tau_{D}}\frac{\partial}{\partial D}\left[(D-\bar{D})P\right]\nonumber\\
&+&\sigma^2\frac{\partial^2}{\partial D^2}\left(DP\right)
\end{eqnarray}
with the initial condition being $P(x,D,0\vert D_0)=\delta(x-x_0)\delta(D-D_0)$ that is
equivalent to
\begin{eqnarray}
\frac{\partial P(x,D,t\vert x_0,D_0)}{\partial t}&=&\frac{1}{\tau_{x}}\frac{\partial}{\partial x}\left[(x-\bar{x})P\right]\nonumber\\
&+&D\frac{\partial^2}{\partial x^2}P-\frac{\partial}{\partial D}J_D\,,
\end{eqnarray}
where $J_D(D,t)$ is the diffusivity flux, i.\,e. 
$J_D(D,t)=-\frac{1}{\tau_{D}}\left[(D-\bar{D})P\right]-\sigma^2\frac{\partial}{\partial D}\left(DP\right)$. The PDF can be translated in the Fourier (space)-Laplace (diffusivity) domain through the integral transform
\begin{eqnarray}
    P^*(q,s,t|x_0,D_0)&=&\int\displaylimits\displaylimits_{-\infty}^\infty dx \int\displaylimits\displaylimits_{0}^\infty dD\, e^{-sD-iqx}\nonumber\\
    &\times&P(x,D,t|x_0,D_0)\
\end{eqnarray}
from which we deduce the new equation:
\begin{eqnarray}
&&\frac{\partial}{\partial t}P^*+Q(s)\frac{\partial}{\partial s}P^*+\frac{1}{\tau_{x}}q\frac{\partial}{\partial q}P^*\\\nonumber
&&=\left(-i\frac{1}{\tau_{x}}\bar{x}q-\frac{1}{\tau_{D}}\bar{D} s\right)P^*+J_D(D=0,t)\,,
\end{eqnarray}
where $Q(s)=\left(\sigma^2s^2+\frac{1}{\tau_{D}}s-q^2\right)$ and $J_D(D=0,t)=(\bar{D}/\tau-\sigma^2)P(q,D=0|x_0,D_0)$.

To ensure diffusivity reaches a stationary distribution, we focus on the case, when there is a reflecting boundary condition at $D=0$. Therefore, the flux cancels at $D=0$ from which $J_D(D=0,t)=0$, and we then obtain
\begin{eqnarray}\label{eq:FourierLaplace_noflux}
\frac{\partial}{\partial t}P^*&+&Q(s)\frac{\partial}{\partial s}P^*+\frac{1}{\tau_{x}}q\frac{\partial}{\partial q}P^*\nonumber\\
&=&\left(-i\frac{1}{\tau_{x}}\bar{x}q-\frac{1}{\tau_{D}}\bar{D} s\right)P^*
\end{eqnarray}
with the initial condition taking the form $P^*(q,s,0\vert x_0,D_0)=e^{-iqx_0}e^{-sD_0}$.


\subsubsection{Method of characteristics}
Our equation (\ref{eq:FourierLaplace_noflux}) is a first order partial differential equation. To solve it, we use the conventional method of characteristics. The Lagrange-Charpit equations \cite{Delgado1997} corresponding to the problem are
\begin{eqnarray}
    dt&=&\frac{dq}{\frac{1}{\tau_{x}}q}=\frac{ds}{\left(\sigma^2s^2+\frac{1}{\tau_{D}}s-q^2\right)}\nonumber\\
    &=&\frac{dP^*}{\left(-i\frac{1}{\tau_{x}}\bar{x}q-\frac{1}{\tau_{D}}\bar{D} s\right)}
\end{eqnarray}
from which we obtain a system of differential equations 
\begin{equation}\label{eq:char_method_system_equation}
\left\{
    \begin{array}{ll}
    \frac{dq}{dt }=\frac{1}{\tau_{x}}q\,,\\\\
\frac{ds}{dt }=\sigma^2s^2+\frac{1}{\tau_{D}}s-q^2\,,\\\\
\frac{dP^*}{dt }=\left(-i\frac{1}{\tau_{x}}\bar{x}q-\frac{1}{\tau_{D}}\bar{D} s\right)P^*\,.
    \end{array}
\right.
\end{equation}

The first equation of the system (\ref{eq:char_method_system_equation}) yields 
\begin{equation}\label{eq:char_method_result_first_eq}
q=C_1e^{\frac{t}{\tau_x}}    
\end{equation}
with $C_1$ an integration constant.

 We then substitute it into the second equation to get
 \begin{equation}\label{carac_sD}
     \frac{ds}{dt }-\sigma^2s^2-\frac{1}{\tau_{D}}s+C_1^2e^{2t/\tau_x}=0\,.
 \end{equation}

After the variable change $s=-\frac{1}{\sigma^2}y'/y$ and the coordinate change $v=\exp(2t/\tau_x)$ we come to the equation
\begin{equation}
v\frac{d^2y}{dv^2}+\left(1-\frac{\tau_{x}}{2\tau_{D}}\right)\frac{dy}{dv}-\frac{\sigma^2 C_1^2\tau_{x}^2}{4}y=0\,.
\end{equation}

\subsubsection{Solving the second equation}
Our equation is of the following form
\begin{equation}
    vy''+(1-a)y'-by=0\,,
\end{equation}
for which the solution is expressed in terms of modified Bessel functions \cite{abr64}, namely
\begin{eqnarray}
    y(v) &=& c_1 b^{a/2} v^{a/2} \Gamma(1 - a) I_{-a}(2 \sqrt{bv})\\\nonumber
    &+& (-1)^{a} c_2 b^{a/2} v^{a/2} \Gamma(a + 1) I_a(2 \sqrt{bv})\,.
\end{eqnarray}
with two integration constants, $c_1$ and $c_2$.
Translating the solution back to our parameters and defining $C_2=c_1/c_2$, as well as $A=c_2(-1)^{\frac{\tau_{x}}{2\tau_{D}}}  \Gamma(\frac{\tau_{x}}{2\tau_{D}} + 1)$, we get
\begin{eqnarray}
    y(t) &=&A  \left(\frac{\sigma^2 C_1^2\tau_{x}^2}{4}\right)^{\frac{\tau_{x}}{4\tau_{D}}}
    e^{t/(2\tau_D)}\nonumber\\
    &&\times
   \Bigg( 
   C_2
    \frac{\Gamma(1 - \frac{\tau_{x}}{2\tau_{D}})}{A} I_{-\frac{\tau_{x}}{2\tau_{D}}}(\sigma\tau_{x} |C_1|e^{t/\tau_x})\nonumber\\
&& \qquad+  I_\frac{\tau_{x}}{2\tau_{D}}(\sigma\tau_{x} |C_1|e^{t/\tau_x})
\Bigg)\,.
\end{eqnarray}

From this we can deduce $s$, i.\,e.
\begin{eqnarray}
s&=&-
\frac{1}{\sigma}|C_1|e^{t/\tau_x}\Big(
C_2\frac{\Gamma(1 - \frac{\tau_{x}}{2\tau_{D}})}{A} I_{1-\frac{\tau_x}{2\tau_D}}(\sigma\tau_x|C_1|e^{t\tau_x})\nonumber\\
&+&I_{-1+\frac{\tau_x}{2\tau_D}}(\sigma\tau_x|C_1|e^{t\tau_x})\Big)\nonumber\\
&\times&\Big(C_2\frac{\Gamma(1 - \frac{\tau_{x}}{2\tau_{D}})}{A} I_{-\frac{\tau_x}{2\tau_D}}(\sigma\tau_x|C_1|e^{t\tau_x})\nonumber\\
&+&I_{\frac{\tau_x}{2\tau_D}}(\sigma\tau_x|C_1|e^{t\tau_x})\Big)^{-1}\,.
\end{eqnarray}
So we obtain
\begin{equation}
    C_2=-\frac{A}
{\Gamma(1 - \frac{\tau_{x}}{2\tau_{D}})}
W\,,
\end{equation}
where
  \begin{equation}
    W=  
\left(
\frac{
I_{-1+\frac{\tau_{x}}{2\tau_{D}}}(\sigma\tau_{x} |q|)
+
\frac{s\sigma}{|q|}
I_{\frac{\tau_{x}}{2\tau_{D}}}(\sigma\tau_{x} |q|)    
}{
\frac{s\sigma}{|q|}
I_{-\frac{\tau_{x}}{2\tau_{D}}}(\sigma\tau_{x} |q|)  
    +
  I_{1-\frac{\tau_{x}}{2\tau_{D}}}(\sigma\tau_{x} |q|)    
}
\right)\,.
\end{equation}

\subsubsection{Third equation}
Then the last equation in Eq.(\ref{eq:char_method_system_equation}) is written as
\begin{equation}
    \int\displaylimits \frac{dP}{P}=-i\frac{1}{\tau_{x}}\bar{x}\int\displaylimits q(t)dt-\frac{\bar{D}}{\tau_{D}}\int\displaylimits s(t)dt\,,
\end{equation}
for which the solution reads
\begin{equation}
    P=C_4\exp\left(-i\frac{1}{\tau_{x}}\bar{x}\int\displaylimits q(t)dt-\frac{\bar{D}}{\tau_{D}}\int\displaylimits s(t)dt\right)\,,
\end{equation}
where $q$ is substituted from Eq.(\ref{eq:char_method_result_first_eq}). By definition we have $\int\displaylimits s(t)dt=-\frac{1}{\sigma^2}\ln (y(t))$  from which we get
\begin{equation}\label{eq:char_meth_P_fct_H}
    P=H(C_1,C_2)\exp\left(-i\bar{x}C_1e^{\frac{t}{\tau_x}}\right) y^{\frac{\bar{D}}{\sigma^2\tau_{D}}}\,,
\end{equation}
where $H(C_1,C_2)$ depends on the constants $C_1$ and $C_2$.
\subsubsection{Initial condition}
At $t=0$, we have $P(q,s,t=0)=e^{-iqx_0-sD_0}$ so that
\begin{eqnarray}\label{eq:char_meth_H}
   H(C_1,C_2)   &=&     \exp\left( -iC_1x_0
   \right)
   \exp\left(i\bar{x}C_1\right)\\\nonumber
&&\times   
   \exp\left(
   -D_0F(C_1,C_2)
   \right)\\\nonumber
&&\times   \left[G(C_1,C_2)\right]^{-\frac{\bar{D}}{\sigma^2\tau_{D}}}\,,
   \end{eqnarray}
where $G(C_1,C_2)=y(C_1,C_2,t=0)$ and $F(C_1,C_2)=s(C_1,C_2,t=0)$.
We then replace $C_1(q,t),C_2(q,s,t)$ by their expressions in $P$.


\subsubsection{Averaging over $D_0$}  
Injecting Eq.(\ref{eq:char_meth_H}) into Eq.(\ref{eq:char_meth_P_fct_H}), we obtain the following propagator
\begin{eqnarray}
    P^*(q,s,t|x_0,D_0)&=&     \exp\left(
   -D_0F(C_1,C_2)
   \right)\\\nonumber
&\times&   
 \exp\left( -iC_1(x_0-\bar{x}(1-e^{\frac{t}{\tau_x}}))
     \right) \\\nonumber
&\times&   
\left(\frac{y}{G(C_1,C_2)}\right)^{\frac{\bar{D}}{\sigma^2\tau_{D}}}\,.
\end{eqnarray}
Next, we integrate over the initial distribution of diffusivity $D_0$ in the form
\begin{equation}
    \Pi(D_0)=\frac{D_0^{\nu-1}}{\Gamma(\nu)\bar{D}^\nu}\exp(- D_0/\bar{D})
\end{equation}
such that the propagator becomes
\begin{eqnarray}\nonumber 
     &&P^*(q,s,t|x_0)=
   \exp\left(-iq\left(\bar{x}(1-e^{-t/\tau_x})+x_0e^{-t/\tau_x}\right)\right)\\
   &&\times
\left(\frac{y(C_1,C_2)}{G(C_1,C_2) \left(1+\bar{D}F(C_1,C_2)\right)}\right)^{\frac{\bar{D}}{\sigma^2\tau_{D}}}\,.
\end{eqnarray}

\subsubsection{Averaging over $D$}
To find the characteristic function $P^*(q,t|x_0)$, we average over $D$ by simply setting 
$s=0$. Thus, we deduce 
  \begin{equation}
    W=  
\left(
\frac{
I_{-1+\frac{\tau_{x}}{2\tau_{D}}}(\sigma\tau_{x} |q|)
}{
  I_{1-\frac{\tau_{x}}{2\tau_{D}}}(\sigma\tau_{x} |q|)    
}
\right)\,.
\end{equation}
In this case we inserted expressions for $y(C_1,C_2)$ and $G(C_1,C_2)$ to get this expression.
Next, we can also include the expression for $F(C_1,C_2)$. Using the property
 $I_{-\nu}(z)=I_{\nu}(z)+(2/\pi)\sin(\nu\pi)K_{\nu}(z)$
 and the Wronskian formula $W\lbrace K_{-\mu}(b),I_{-\mu}(b)\rbrace=I_{-\mu}(b)K_{1-\mu}(b)+K_{-\mu}(b)I_{1-\mu}(b)=1/b$, we come to the full characteristic function Eq.(\ref{Eq:char_fun_marginal}).
One can check normalization by setting $q=0$ and verify that $P^*(q= 0,t|x_0)=1$.

\section{Large order development of $I_\beta(b)$}
\label{appendix:BesselI_beta_large_order}
Based on the asymptotic large order expansion for $I_{\mu-1}(b)$, we start with the integral representation
  \begin{equation}
      I_{\beta}(b)=\frac{(b/2)^{\beta}}{\sqrt{\pi}\Gamma(\beta+1/2)}\int\displaylimits_{-1}^1(1-t^2)^{\beta-1/2}e^{-bt}dt\,.
  \end{equation}
With help of the variable change $-u^2=\ln(1-t^2)$ from which $t=\sqrt{1-e^{-u^2}}$ and $dt=\frac{u e^{-u^2}}{\sqrt{1-e^{-u^2}}}du$ we proceed to
\begin{equation}
      I_{\beta}(b)=\frac{(b/2)^{\beta}}{\sqrt{\pi}\Gamma(\beta+1/2)}\int\displaylimits_{\mathbb{R}}e^{-\beta u^2}f(u)du\,,
  \end{equation}
 where $f(u)=ue^{-b\sqrt{1-e^{-u^2}}}\frac{ e^{-u^2/2}}{\sqrt{1-e^{-u^2}}}$. Next, we find the asymptotic behavior for large $\mu$ written as
\begin{equation}
   \frac{z^\beta}{ u^\beta\Gamma(\beta+1/2)}=\frac{1}{\sqrt{2\pi}}\left(\frac{eb}{2\beta}\right)^\beta+O(\beta^{-1-\beta})\,.
\end{equation}
 For $f(u)$ we expand $u$ for small order
 \begin{equation}
     f(u)\approx e^{-bu} (1-u^2/2)\,.
 \end{equation}

The integral over $u$ yields
\begin{equation}
   \int\displaylimits_\mathbb{R}e^{-\beta u^2}f(u)du= e^{b^2/(4\beta)}\frac{\sqrt{\pi}}{4\beta^{3/2}}\left(\frac{8\beta^2-b^2}{2\beta}-1\right)\,.
\end{equation}

So we deduce
\begin{equation}
     I_{\beta}(b)\approx\frac{e^{b^2/(4\beta)}}{\sqrt{2}}\left(\frac{eb}{2\beta}\right)^\beta \frac{1}{4\beta^{3/2}}\left(\frac{8\beta^2-b^2}{2\beta}-1\right)\,.
\end{equation}

\section{Fourth moment computation}\label{Appendix:fourth_moment}
In this section we study the fourth moment in the case of temperature fluctuations. For this purpose we proceed to the variable change $z_t=x_t^2$
for which the integral representation is
\begin{eqnarray}
z_t&=&z_0e^{-2t/\tau_x}+\int\displaylimits_0^t2D_s e^{2(s-t)/\tau_x}ds\nonumber\\
&&+\int\displaylimits_0^t2 e^{2(s-t)/\tau_x}\sqrt{2z_sD_s}dW_s
\end{eqnarray}
from which we deduce the integral representation for the fourth moment
\begin{eqnarray}\nonumber
\langle x_t^4\rangle &=& \langle z_t^2\rangle=\langle x_0^4\rangle e^{-4t/\tau_x}\\\nonumber
   &+&4e^{-4t/\tau_x}\int\displaylimits_0^t\int\displaylimits_0^te^{2(s_1+s_2)/\tau_x}\langle D_{s_1}D_{s_2}\rangle ds_1ds_2\\
   &+&8\int\displaylimits_0^te^{4(s-t)/\tau_x}
\langle x_{s}^2D_{s}\rangle ds\,.
\end{eqnarray}
We proceed to use the It$\hat{\rm o}$ formula for the variable change from $x$ to $x^2$
\begin{equation}
\left\{
    \begin{array}{ll}
dx_t^2=\left(-\frac{2}{\tau_{x}}x_t^2+2D_t\right)dt+2x_t\sqrt{2D_t}d W_t^{(1)} \\
dD_t=-\frac{1}{\tau_{D}}(D_t-\bar{D})dt+\sigma\sqrt{2D_t}dW_t^{(2)}
    \end{array}\,.
\right.
\end{equation}
From the It$\hat{\rm o}$ product rule we have
\begin{eqnarray}
    x_t^2D_t&=&x_0^2D_0+\int_0^t x_s^2dD_s\nonumber\\
    &+&\int_0^tD_sdx_s^2+\int_0^t \langle dx_s^2dD_s\rangle
\end{eqnarray}
with the last term being equal to zero because of the independence of Wiener processes $W_t^{(1)}$ and $W_t^{(2)}$. From $\langle x_t^2D_t\rangle$ it follows an integral equation in the form
\begin{equation}
    y(t)=y(0)-a\int_0^ty(s)ds+\int_0^t b(s)ds\,,
\end{equation}
where $a=\frac{2}{\tau_x}+\frac{1}{\tau_D}$ and $b(t)=\langle x_s^2\rangle\bar{D}/\tau_D+2\langle D_s^2\rangle$. Taking the derivative on both sides, we get
\begin{equation}
    y'(t)=-ay(t)+b'(t)\,,
\end{equation}
from which we have
\begin{equation}
     \langle x_t^2D_t\rangle=\int_0^t e^{(s-t)/\tau_x}\left(\langle x_s^2\rangle\bar{D}/\tau_D+2\langle D_s^2\rangle\right)ds\,.
     \end{equation}
 The exact calculation is achievable, although it is tedious and cumbersome. Instead we find it more informative to focus on the long-time limit. When $t\to\infty$, only constant terms contribute to convolution integrals from which we deduce Eq. (\ref{eq:fourth_mom_long_time}).


\end{document}